\documentclass[a4paper]{jpconf}
\usepackage{lmodern}
\usepackage{amssymb,amsfonts,amsmath,,amscd}
\newcommand{\be}{\begin{equation}}
\newcommand{\ee}{\end{equation}}
\newcommand{\bea}{\begin{eqnarray}}
\newcommand{\eea}{\end{eqnarray}}
\newcommand{\nn}{\nonumber \\}

\begin{document}
\title{Long ${\cal N}=2$, $4$ multiplets in supersymmetric mechanics}

\author{Evgeny Ivanov$^1$, Antonio Rivasplata Mendoza$^{1,2}$, Stepan Sidorov$^1$}

\address{$^1$ Bogoliubov Laboratory of Theoretical Physics, JINR, 141980 Dubna, Moscow Region, Russia}
\address{$^2$ Departamento de F\'isica Universidad Nacional de Trujillo, Trujillo, Per\'u}

\ead{eivanov@theor.jinr.ru, antrivas@unitru.edu.pe, sidorovstepan88@gmail.com}

\begin{abstract}
We define $SU(2|1)$ supermultiplets described by chiral superfields having non-zero external spins with respect to $SU(2) \subset SU(2|1)$ and show that their splitting into ${\cal N}=2$, $d=1$ multiplets contains the so called ``long'' indecomposable ${\cal N}=2$, $d=1$ multiplets ${\bf (2, 4, 2)}_{\rm l}$. We give superfield formulation for this type of ${\cal N}=2$ long multiplets and construct their most general superfield action. A simple example of long ${\cal N}=4$, $d=1$ multiplet is also considered, both in the superfield and the component formulations.
\end{abstract}

\section{Introduction}
In \cite{DSQM}, $SU(2|1)$ supersymmetric mechanics was proposed as a deformation of the standard ${\cal N}=4$ mechanics
by a mass parameter $m$\,. Superfield approach based on the deformed $SU(2|1)$ superspaces allowed to reproduce many previously
known models \cite{WS,BelNer,R1,R2} and to construct new ones \cite{DSQM,SKO,ISTconf,DHSS}.
In the paper \cite{Casimir}, $SU(2|1)$ supersymmetric quantum mechanics was obtained via dimensional reduction from
the superconformal model on the four-dimensional curved space-time $S^3 \times \mathbb{R}$ and applied to compute vacuum energy of the model.
For simplicity, the authors considered supersymmetric mechanics in the framework of ${\cal N}=2$, $d=1$ supersymmetry and
revealed a new type of supermultiplets, the so-called ``long multiplets''.
As was shown in \cite{LM}, the long ${\cal N}=2$ multiplet can be embedded into a generalized $SU(2|1)$ chiral multiplet described
by a chiral superfield $\Phi_A$ carrying some external index $A$ with respect to the subgroup $SU(2)$ of the supergroup $SU(2|1)$.

Generalizations to ${\cal N}=4$ supersymmetry with various extended sets of component fields were considered in \cite{GonTo1,GonTo,Katona}.
The main distinguishing feature  of long (non-minimal) multiplets is that they accommodate extended sets of component fields.
The long ${\cal N}=2$ multiplet \cite{Casimir} can be interpreted as a deformation of the pair of chiral multiplets ${\bf (2, 2, 0)}$
and ${\bf (0, 2, 2)}$ by a mass-dimension parameter, {\it i.e.} it has an extended set of component fields ${\bf (2, 4, 2)}_{\rm l}$\,.
The long multiplet ${\bf (4, 8, 4)}_{\rm l}$ considered in \cite{Katona} joins two ${\cal N}=4$ chiral multiplets ${\bf (2, 4, 2)}$
through a dimensionless parameter.

In this contribution we give a brief account of the long ${\cal N}=2$ multiplet, as it was discussed in \cite{LM}, and present
some new results for the long ${\cal N}=4$ multiplet suggested in \cite{Katona}.
To be more precise, we give the superfield description for the long ${\cal N}=2$, $4$ multiplets which were studied at the component
level in \cite{Casimir,Katona}.

\section{$SU(2|1)$ supersymmetric mechanics}

We proceed from the centrally-extended superalgebra $\hat{su}(2|1)$  with the following non-vanishing (anti)commutators:
\bea
    &&\lbrace Q^{i}, \bar{Q}_{j}\rbrace = 2m\left( I^i_j -\delta^i_j F\right)+ 2\delta^i_j H,\qquad\left[I^i_j,  I^k_l\right]
    = \delta^k_j I^i_l - \delta^i_l I^k_j,\nn
    &&\left[I^i_j, \bar{Q}_{l}\right] = \frac{1}{2}\,\delta^i_j\bar{Q}_{l}-\delta^i_l\bar{Q}_{j},\qquad \left[I^i_j, Q^{k}\right]
    = \delta^k_j Q^{i} - \frac{1}{2}\,\delta^i_j Q^{k},\nn
    &&\left[F, \bar{Q}_{l}\right]=-\frac{1}{2}\,\bar{Q}_{l},\qquad \left[F, Q^{k}\right]=\frac{1}{2}\,Q^{k}.\label{algebra}
\eea
Its bosonic sector contains the central charge generator $H$ (commuting with all other generators) and
the $U(2)_{\rm int}$ generators $I^i_j$ and $F$.
In the limit $m=0$, this superalgebra becomes the standard ${\cal N}=4$, $d=1$ Poincar\'e superalgebra.

The supersymmetric $SU(2|1)$ transformations of the superspace coordinates $\zeta := \left\lbrace t,\theta_i,\bar{\theta}^k\right\rbrace$,
$\bar{\theta}^i=\overline{\left(\theta_i\right),}$  are given by
\bea
    \delta\theta_{i}=\epsilon_{i} +
    2m\,\bar{\epsilon}^k\theta_k\theta_{i},\quad
    \delta \bar{\theta}^{i}=\bar{\epsilon}^{i} -
    2m\,\epsilon_k\bar{\theta}^k\bar{\theta}^{i},\qquad
    \delta t = i\left(\bar{\epsilon}^k\theta_k + \epsilon_k\bar{\theta}^k\right).
\eea
The $SU(2|1)$ measure invariant under these transformations is
\bea
    d\zeta = dt\,d^2\theta\,d^2\bar{\theta}\left(1+2m\,\bar{\theta}^k\theta_k\right),\quad \delta\left(d\zeta\right)=0.
\eea
The left chiral subspace $\zeta_L=\left\lbrace t_L,\theta_i\right\rbrace\,,$  where $t_L$ is defined as
\bea
    t_L =t+i\bar{\theta}^k\theta_k - \frac{i}{2}\,m \left(\theta\right)^2\left(\bar\theta\,\right)^2,
\eea
is closed under the $SU(2|1)$ transformations
\bea
    \delta\theta_{i}=\epsilon_{i} +
    2m\,\bar{\epsilon}^k\theta_k\theta_{i},\qquad
    \delta t_L = 2i\bar{\epsilon}^k\theta_k\,.
\eea
Conjugating the coordinates of the subspace $\zeta_L$, one obtains the right-chiral subspace $\zeta_R$.

The $SU(2|1)$ covariant derivatives are defined as
\bea
    {\cal D}^i &=& \left[1+m\,\bar{\theta}^k\theta_k
    -\frac{3m^2}{8}\left(\theta\right)^2\left(\bar{\theta}\,\right)^2\right]\frac{\partial}{\partial\theta_i}
    - m\,\bar{\theta}^i\theta_j\frac{\partial}{\partial\theta_j}-i\bar{\theta}^i \partial_t\nn
    &&+\,m\,\bar{\theta}^i \tilde{F}- m\,\bar{\theta}^j\left(1-m\,\bar{\theta}^k\theta_k \right)\tilde{I}^i_j,\nn
    \bar{{\cal D}}_j &=& -\left[1+ m\,\bar{\theta}^k\theta_k
    -\frac{3m^2}{8}\left(\theta\right)^2\left(\bar{\theta}\,\right)^2\right]\frac{\partial}{\partial\bar{\theta}^j}
    + m\,\bar{\theta}^k\theta_j\frac{\partial}{\partial\bar{\theta}^k}+i\theta_j\partial_t\nn
    &&-\,m\,\theta_j\tilde{F}+m\,\theta_l\left(1-m\,\bar{\theta}^k\theta_k \right)\tilde{I}^l_j,
\eea
where $\tilde{F}$ and $\tilde{I}^i_k$ are the ``matrix'' parts of the generators $F$ and $I^i_k$.
The latter non-trivially act on the covariant derivatives:
\bea
   &&\tilde{I}^i_j \bar{{\cal D}}_{l} = \delta^i_l\bar{{\cal D}}_{j}-\frac{1}{2}\,\delta^i_j\bar{{\cal D}}_{l}\, ,\qquad
   \tilde{I}^i_j {\cal D}^{k} =  \frac{1}{2}\,\delta^i_j {\cal D}^{k} -\delta^k_j {\cal D}^{i} ,\nn
   &&\tilde{F}\bar{{\cal D}}_{l}=\frac{1}{2}\,\bar{{\cal D}}_{l}\,,\qquad \tilde{F} {\cal D}^{k}=-\,\frac{1}{2}\,{\cal D}^{k}.
\eea
An $SU(2|1)$ superfield $\Phi_{A}$ can carry an external $U(2)$ representation corresponding to these matrix parts
and it transforms according to this representations as
\bea
    &&\delta \Phi_{A} = \left(i\delta \hat{h}\tilde{F} - i\delta h_{ij} \tilde{I}^{ij}\right)\Phi_{A}\,,\nn
    &&\delta \hat{h}=-\,im\left(\epsilon_k\bar{\theta}^k+\bar{\epsilon}^k\theta_k\right),\quad
    \delta h_{ij} = im\left(\epsilon_{(i}\bar{\theta}_{j)} +\bar{\epsilon}_{(i} \theta_{j)}\right)
\left(1-m\,\bar\theta^k\theta_k\right).\label{SFtr}
\eea
\subsection{Chiral superfields}
Chiral  $SU(2|1)$ superfields can carry non-zero external spins $s$
with respect to  $SU(2) \subset SU(2|1)$. The simplest chiral superfield with  $s=0$ has the field contents $({\bf 2, 4, 2})$.
As compared to the  $SU(2)$ singlet chiral superfields, the number of component fields in the superfield  $\Phi_A$ carrying
non-zero external spins  $s=1/2, 1, \ldots $ increases according to
    \be
        {\bf\left(2[2s+1], 4[2s+1], 2[2s+1]\right)}.
    \ee
The decomposition of the  $s=0$ chiral supermultiplet $({\bf 2, 4, 2})$ into ${\cal N}=2$ multiplets
is given by a direct sum of two chiral multiplets,  $({\bf 2, 2, 0})$ and $({\bf 0, 2, 2})\,$.
Below, we will consider the analogous decompositions of the  $s\neq 0$ chiral multiplets.

The singlet ($s=0$) chiral superfield $\Phi$ in the $U(2)$ representation $\left(0, 2\kappa\right)$ satisfies the chirality condition
\bea
   \bar{\cal D}_i \Phi = 0,\qquad \tilde{F}\Phi = 2\kappa\,\Phi ,\qquad \tilde{I}^{k}_{l}\Phi = 0.
\eea
In the case of $s=1/2,1,3/2\ldots$, the chiral superfield $\Phi_{(i_1\,\ldots \,i_{2s})}$ belongs
to the $U(2)$ representation $\left(s,2\kappa\right)$ and is defined by the constraints
\bea
    &&\bar{{\cal D}}_j\Phi_{(i_1\,\ldots\, i_{2s})} = 0,\qquad  \tilde{F}\Phi_{(i_1\,\ldots\, i_{2s})} = 2\kappa\,\Phi_{(i_1\,\ldots\, i_{2s})}\,,\nn
    &&\tilde{I}^{k}_{l}\Phi_{(i_1\,\ldots\, i_{2s})} = \sum_{n = 1}^{2s}\left[\delta^k_{i_n}\Phi_{(i_1\,\ldots\, i_{n-1}\,l\,i_{n+1}\,\ldots\, i_{2s})}
    - \frac{1}{2}\,\delta^k_l\Phi_{(i_1\,\ldots\, i_{2s})}\right].
\eea
$SU(2|1)$ transformations of chiral superfields can be found from \eqref{SFtr}.

\subsection{The case $s=1/2$}
The  $SU(2|1)$ chiral superfield  $\Phi_i$ ($i=1,2$) in the  $U(2)$ representation  $\left(1/2 , 2\kappa\right)$ is defined by the constraints
\bea
    \bar{{\cal D}}_j\Phi_i = 0,\qquad \tilde{I}^{k}_{l}\Phi_i = \delta^k_i\Phi_l
    - \frac{1}{2}\,\delta^k_l\Phi_i ,\qquad \tilde{F}\Phi_i = 2\kappa\,\Phi_i .
\eea
The chirality condition is solved by
\bea
    \Phi_i\left(t_L,\theta_i,\bar{\theta}_k\right) &=& \left(1 + 2m\,\bar{\theta}^l\theta_l\right)^{-\kappa}
    \left[1 - \frac{3m^2}{16}\left(\theta\right)^2\left(\bar{\theta}\,\right)^2\right]\phi_{i}\left(t_L,\theta_i\right)\nn
    &&-\,m\left(\frac{1}{2}\,\delta^j_i\,\bar{\theta}^k\theta_k -\bar{\theta}^j\theta_i\right)\phi_j\left(t_L,\theta_i\right),\nn
    \phi_{i}\left(t_L,\theta_i\right) &=& z_i + \theta_i\psi - \sqrt{2}\,\theta^k\psi_{(ik)}+ \theta_k\theta^k B_i\,.
\eea
The superfields $\Phi_i$ and $\phi_i$ transform as
\bea
    \delta\Phi_i &=& m\left(1 - m\,\bar{\theta}^l\theta_l\right)\left[\frac{1}{2}\,\delta^j_i
    \left(\epsilon_k\bar{\theta}^k+\bar{\epsilon}^k\theta_k\right)
    -\left(\epsilon_i\bar{\theta}^j+\bar{\epsilon}^j\theta_i\right)\right]\Phi_j
    + 2\kappa m \left(\epsilon_k\bar{\theta}^k +\bar{\epsilon}^k\theta_k\right) \Phi_i,\nn
    \delta\phi_i &=& 4\kappa m\left(\bar{\epsilon}^k\theta_k\right)\phi_i + 2m\left(\frac{1}{2}\,\delta^j_i\,\bar{\epsilon}^k\theta_k
    - \bar{\epsilon}^j\theta_i\right)\phi_j\, .
\eea
The relevant $SU(2|1)$ transformations of the component fields read
\bea
    &&\delta z^i  =-\,\epsilon^i\psi -\sqrt{2}\,\epsilon_k\psi^{(ik)},\qquad\delta B^i = -\,\bar{\epsilon}^i\left(i\,\nabla_t\psi - \frac{m}{2}\,\psi\right)-\sqrt{2}\,\bar{\epsilon}_k
    \left(i\,\nabla_t\psi^{(ik)}+\frac{3m}{2}\,\psi^{(ik)}\right),\nn
    &&\delta \psi =  \bar{\epsilon}^k\left(i\,\nabla_t z_k + \frac{3m}{2}\,z_k\right)-\epsilon^k B_k ,\qquad
    \delta \psi^{(ik)} = \sqrt{2}\, \bar{\epsilon}^{(k}\left[i\,\nabla_t z^{i)} - \frac{m}{2}\,z^{i)}\right]
    -\sqrt{2}\,\epsilon^{(i} B^{k)},\label{SU21tr}
\eea
where
\bea
    \nabla_t = \partial_t +2 i\kappa m\,, \qquad \bar{\nabla}_t = \partial_t - 2 i\kappa m\,.
\eea

\subsection{Decomposition into ${\cal N}=2$ multiplets}
Singling out the subset of ${\cal N}=2$ transformations associated with the parameter $\epsilon_1\equiv\epsilon$ in \eqref{SU21tr},
we can identify the component fields $\left(z_i,\psi^{(ik)},\psi, B^i\right)$ with the system of three ${\cal N}=2$ multiplets:
\be
    {\rm one\;long\;multiplet}\;{\bf (2, 4, 2)}_{\rm l}\quad \oplus \quad {\rm one\;short\;multiplet} \;({\bf 2, 2, 0})
\quad \oplus \quad {\rm one\;short\;multiplet} \; {\bf (0, 2, 2)}.\nonumber
\ee
The ${\cal N}=2$-irreducible multiplets $({\bf 2, 2, 0})$ and $({\bf 0, 2, 2})$ are composed of the fields $(z_2, \psi^{(11)})$
and $( \psi^{(22)},B^2)$, while the rest of component fields $\left(z_1,\psi^{(12)},\psi, B^1\right)$ forms
a multiplet with the field contents $({\bf 2, 4, 2})_{\rm l}$ that was called ``long'' multiplet \cite{Casimir}.
In the limit $m=0$, the indecomposable long ${\cal N}=2$ multiplet $({\bf 2, 4, 2})_{\rm l}$ splits into the direct sum
of two ``short'' irreducible ${\cal N}=2$ multiplets ${\bf (2, 2, 0)}$ and ${\bf (0, 2, 2)}$.
At $m \neq 0$, such a splitting cannot be accomplished by any field redefinition.

In the general case $s>0$, we have the following
sum of ${\cal N}=2$ multiplets:
\be
    2s \;{\rm long\;multiplets}\quad \oplus \quad {\rm one \; short\;multiplet} \;({\bf 2, 2, 0})\quad \oplus \quad {\rm one\; short\;multiplet}
\; ({\bf 0, 2, 2}) .\nonumber
\ee
{}From this decomposition one can figure out that all long multiplets have mass-dimension parameters proportional to $m$\,.

The relevant ${\cal N}=2$ subalgebra of \eqref{algebra} can be identified with
\bea
    \lbrace Q, \bar{Q}\rbrace = 2\left(H - \Sigma\right), \quad Q^2 = \bar Q^2 = 0\,,
\eea
where
\bea
    Q\equiv Q^1,\qquad \bar{Q}\equiv\bar{Q}_1,\qquad \Sigma \equiv m\left(F-I^1_1\right).
\eea
If we forget about the $su(2|1)$ origin of this superalgebra, the presence of the central $\Sigma$ is not necessary
since it can always be removed by a field redefinition, and $H -\Sigma$ can be chosen as the Hamiltonian:
\bea
    H -\Sigma\;\; \rightarrow\;\; H\equiv i\partial_t\,.
\eea
For the long multiplet ${\bf (2, 4, 2)}_{\rm l}$\,, this shift can be performed through the redefinitions
\bea
    && z=z_1e^{i\left(2\kappa-1/2\right)mt},\qquad \xi=\left(\frac{\psi}{\sqrt{2}}-\psi_{(12)}\right)e^{i\left(2\kappa-1/2\right)mt},\nn
    && B=-B^1e^{i\left(2\kappa-1/2\right)mt},\qquad \pi=\left(\frac{\psi}{\sqrt{2}}+\psi_{(12)}\right)e^{i\left(2\kappa-1/2\right)mt}.\label{redef}
\eea
Then, the corresponding ${\cal N}=2$ supersymmetry transformations,
\bea
    &&\delta z  = -\,\sqrt{2}\,\epsilon\,\xi,\qquad\delta \xi =  \sqrt{2}\, i\bar{\epsilon}\dot{z},\nn
    &&\delta \pi = -\,\sqrt{2}\,\epsilon B + \underline{\sqrt{2}\,m\,\bar{\epsilon}z},\qquad\delta B =
    \sqrt{2}\,i\bar{\epsilon}\,\dot{\pi}-\underline{\sqrt{2}\,m\,\bar\epsilon\,\xi},\label{N2tr}
\eea
close on the standard ${\cal N}=2$ supersymmetry algebra
\bea
    \left\lbrace Q, \bar{Q}\right\rbrace = 2H,  \qquad \left\lbrace Q, Q\right\rbrace =
    \left\lbrace \bar Q, \bar Q\right\rbrace = 0,\quad H=i\partial_t \,.\label{N2algebra}
\eea

The free $SU(2|1)$ Lagrangian is given by
\bea
    {\cal L}^{\rm free} = \frac{1}{4}\int d^2\theta\,d^2\bar{\theta}\left(1+2m\,\bar{\theta}^k\theta_k\right)\Phi_i \bar{\Phi}^i.
\eea
Rewriting it in terms of the component fields,
\bea
    {\cal L}^{\rm free}&=& \bar{\nabla}_t {\bar{z}}^i\,\nabla_t {z}_i+ \frac{i}{2}\left(\bar{\psi}\,\nabla_t\psi
    + \psi\,\bar{\nabla}_t\bar{\psi} \right)+\frac{i}{2}\left(\bar{\psi}_{(ik)}\,\nabla_t\psi^{(ik)} + \psi^{(ik)}\,\bar{\nabla}_t\bar{\psi}_{(ik)}\right)+B^i\bar{B}_i\nn
    && -\,\frac{i}{2}\,m\left(z_i\bar{\nabla}_t {\bar{z}}^i- \bar{z}^i\nabla_t {z}_i\right) - \frac{3m^2}{4}\,z_i\bar{z}^i
     +\,\frac{m}{2}\left(\psi\bar{\psi}-3\psi^{(ik)}\bar{\psi}_{(ik)}\right),\label{SU21free}
\eea
one can check that it splits into a sum of the three free ${\cal N}=2$ Lagrangians:
\be
{\cal L}^{\rm free}={\cal L}^{\rm free}_{({\bf 2, 4, 2})_{\rm l}}
+{\cal L}^{\rm free}_{\bf( 0, 2, 2)}+{\cal L}^{\rm free}_{\bf( 2, 2, 0)}\,.\label{N2free}
\ee
After the redefinition \eqref{redef}, the component Lagrangian of the long multiplet in \eqref{N2free} reads
\bea
    {\cal L}^{\rm free}_{({\bf 2, 4, 2})_{\rm l}} = \dot{\bar{z}}\dot{z}+ \frac{i}{2}\left(\bar{\xi}\dot{\xi} - \dot{\bar{\xi}}\xi\right)
    +\frac{i}{2}\left(\bar{\pi}\dot{\pi} - \dot{\bar{\pi}}\pi\right) +B\bar{B}-\underline{m\left(\xi\bar{\pi}+\pi\bar{\xi}\,\right)}
    -\underline{m^2 z\bar{z}}.\label{lN2free}
\eea

\section{The long ${\cal N}=2$ multiplet}
Now we consider a superfield description for the long ${\cal N}=2$ supermultiplet defined by the transformations \eqref{N2tr}.
First we define  ${\cal N}=2$ covariant derivatives $D$, $\bar{D}\,,$
\bea
    D=\frac{\partial}{\partial\theta}-i\bar{\theta}\partial_t\, ,\qquad
    \bar{D}=-\,\frac{\partial}{\partial \bar{\theta}}+i\theta\partial_t\, ,\qquad
    \left\lbrace D, \bar{D}\right\rbrace = 2i\partial_t \,.
\eea
The  ${\cal N}=2$,  $d=1$ superspace coordinates  $\left\lbrace t,\theta,\bar{\theta}\right\rbrace$ transform in the familiar way:
\bea
    \delta\theta =\epsilon, \qquad \delta\bar{\theta}=\bar{\epsilon},\qquad
    \delta t=i\left(\epsilon\,\bar\theta + \bar{\epsilon}\,\theta\right).
\eea

The chiral superfields are defined by the standard conditions
\bea
      \bar{D} Z = 0,\qquad \bar{D} \Pi = 0.
\eea
The bosonic superfield  $Z$ describes an irreducible multiplet $({\bf 2,2,0})$,
while the fermionic superfield  $\Pi$ has the field contents $({\bf 0, 2, 2})$. The component expansion of  $\Pi$ and  $Z$ reads
\bea
    Z=z + \sqrt{2}\,\theta\,\xi
    - i\theta\bar{\theta}\dot{z},\qquad
    \Pi = \pi +
    \sqrt{2}\,\theta B - i\theta\bar{\theta}\,\dot{\pi}.
\eea
The ``passive'' superfield transformations  $\delta\Pi = \delta Z =0$ amount to the two independent sets of transformations
for the component fields:
\bea
     \delta z  = -\,\sqrt{2}\,\epsilon\,\xi,\quad\delta \xi =  \sqrt{2}\, i\bar{\epsilon}\dot{z},\qquad
     \delta \pi = -\,\sqrt{2}\,\epsilon B,\quad\delta B =
    \sqrt{2}\,i\bar{\epsilon}\,\dot{\pi}.\label{m0tr}
\eea

The long multiplet is described by the pair of fermionic and bosonic ${\cal N}=2$ superfields  $\Psi$ and  $Z$ (of the same dimension as $\Pi$ and $Z$ above)
which are subjected to the following conditions, with $m$ being a deformation parameter:
\bea
     \bar{D} \Psi = \sqrt{2}\,m Z,\qquad \bar{D} Z = 0. \label{LongConstr}
\eea
As a solution of (\ref{LongConstr}) (non-singular at $m=0$), the superfield $\Psi$ can be represented as
\be
    \Psi = \Pi - \sqrt{2}\,m\,\bar\theta Z,\qquad \bar{D} \Pi = 0.
\ee
The first condition in (\ref{LongConstr}) expresses some components of $\Psi$ through the components of $Z$
and so forces the superfunction $\Pi$ to transform through $Z$.

The transformations $\delta\Psi = \delta Z =0$ give rise to the following transformation law for $\Pi$:
\bea
    \delta\Pi = \sqrt{2}\,m\,\bar{\epsilon}Z. \label{PiZ}
\eea
It generates deformed ${\cal N}=2$ supersymmetry  transformations which coincide with the transformations \eqref{N2tr}.
Thus the considered  multiplet involves an irreducible chiral multiplet ${\bf (2, 2, 0)}$ and a set of fields ${\bf (0, 2, 2)}$ which are described
by the chiral  $d=1$ superfunctions  $Z$ and  $\Pi$, respectively. The quantity  $Z$ is a chiral superfield, while $\Pi$ has
the non-standard transformation law (\ref{PiZ}) $\sim m\,$.
Thus the parameter  $m$ is a deformation parameter responsible for unifying the two former chiral ``short'' multiplets into a single ``long'' multiplet.

\subsection{Invariant Lagrangians}
The most general Lagrangian of the long multiplet can be written down as
\bea
    {\cal L}_{(\Psi,Z)} = \int d\bar{\theta}\, d\theta\left[\bar{D}\bar{Z}\,DZ\,h_0\left(Z,\bar{Z}\right)
    +\Psi\bar{\Psi}\,h_1\left(Z,\bar{Z}\right)+\mu \,h_{(\mu)}\left(Z,\bar{Z}\right)\right],
\eea
where $h_0$, $h_1$, $h_{(\mu)}$ are arbitrary functions and $\mu$ is a mass-dimension parameter.
The free Lagrangian of the long multiplet is given by a sum of the three superfield invariants:
\bea
    {\cal L}^{\rm free}_{({\bf 2, 4, 2})_{\rm l}}=\frac{1}{4}\int d\bar{\theta}\, d\theta\left(\bar{D}\bar{Z}\, DZ
    + 2\Psi\bar{\Psi}-2\mu Z\bar{Z}\right).
\eea
In the component form it reads
\bea
    {\cal L}^{\rm free}_{({\bf 2, 4, 2})_{\rm l}} &=& \dot{\bar{z}}\dot{z}+ \frac{i}{2}\left(\bar{\xi}\dot{\xi} - \dot{\bar{\xi}}\xi\right)
    +\frac{i}{2}\left(\bar{\pi}\dot{\pi} - \dot{\bar{\pi}}\pi\right) +B\bar{B}-m\left(\xi\bar{\pi}+\pi\bar{\xi}\,\right)
    -m^2 z\bar{z}\nn
    &&-\,\mu\left[\frac{i}{2}\left(z\dot{\bar z}-\bar{z}\dot{z}\right)+\xi\bar{\xi}\right].
\eea
The free Lagrangian \eqref{lN2free} corresponds to the choice $\mu = 0$.

\section{The long ${\cal N}=4$ multiplet}
In this section, we give a superfield description of the indecomposable long ${\cal N}=4$ supermultiplet suggested in \cite{Katona}.
The standard ${\cal N}=4$ supersymmetry superalgebra is formed by the (anti)commutators
\bea
    \left\lbrace Q^i, \bar{Q}_j\right\rbrace = 2\delta^i_j H,   \qquad \left\lbrace Q^i, Q^j\right\rbrace =
    \left\lbrace \bar{Q}_i, \bar{Q}_j\right\rbrace =0,\quad H=i\partial_t \,.
\eea
The covariant ${\cal N}=4,\, d=1$ derivatives are defined in the standard way as
\bea
    D^i = \frac{\partial}{\partial\theta_i}-i\bar{\theta}^i\partial_t\,,\quad
    \bar{D}_j = -\,\frac{\partial}{\partial\bar{\theta}^j}+i\theta_j\partial_t\,,\qquad
    \left\lbrace D^i, \bar{D}_j\right\rbrace = 2\delta^i_j H .
\eea
The  ${\cal N}=4$,  $d=1$ superspace coordinates  $\left\lbrace t,\theta_i,\bar{\theta}^j\right\rbrace$ undergo
the transformations:
\bea
    \delta\theta_i =\epsilon_i, \qquad \delta\bar{\theta}^j = \bar{\epsilon}^j,\qquad
    \delta t=i\left(\epsilon_i\bar{\theta}^i + \bar{\epsilon}^i\theta_i\right).
\eea

The indecomposable long ${\cal N}=4$ supermultiplet is parametrized by a real dimensionless parameter $\alpha$ and
is described by the system of complex ${\cal N}=4$ superfields $V$ and $W$ obeying the constraints
\bea
     \bar{D}_i V = i\alpha D_i W,\qquad \bar{D}_i W = 0.\label{alpha}
\eea
In the limit $\alpha=0$, these constraints are reduced to those defining two ordinary $({\bf 2, 4, 2})$ chiral multiplets.
The constraints \eqref{alpha} are solved by
\bea
    &&V\left(t,\theta_i,\bar{\theta}^j\right) = V_0\left(t,\theta_i,\bar{\theta}^j\right) +
    i\alpha\,\bar{\theta}_k\frac{\partial}{\partial \theta_k}\,W\left(t,\theta_i,\bar{\theta}^j\right) ,
    \qquad \bar{D}_k V_0 = 0,\label{solution}
\eea
implying the following transformation properties for the involved objects
\bea
    \delta V_0\left(t_L,\theta_i\right) =  -\,i\alpha\,\bar{\epsilon}_k\frac{\partial}{\partial \theta_k}\,W\left(t_L,\theta_i\right),\qquad
\delta W = \delta V = 0.\label{V0tr}
\eea
In components, the solution \eqref{solution} reads
\bea
    V &=& y + \sqrt{2}\,\theta_i\xi^i +\theta_i\theta^i A + i\bar{\theta}^i\theta_i \dot{y}
+ \sqrt{2}\,i\bar{\theta}^j\theta_j\theta_i\dot{\xi}^i - \frac{1}{4}\,\bar{\theta}^j\bar{\theta}_j\theta_i\theta^i \ddot{y}\nn
    && + i\alpha\,\bar{\theta}_i\left(\sqrt{2}\,\psi^i +2\theta^i C -i\bar{\theta}^i\dot{x}
+ \sqrt{2}\,i\theta^i\bar{\theta}_j\dot{\psi}^j\right) ,\nn
    W &=& x + \sqrt{2}\,\theta_i\psi^i +\theta_i\theta^i C + i\bar{\theta}^i\theta_i \dot{x}
+ \sqrt{2}\,i\bar{\theta}^j\theta_j\theta_i\dot{\psi}^i - \frac{1}{4}\,\bar{\theta}^j\bar{\theta}_j\theta_i\theta^i \ddot{x},
\eea
with
\bea
    &&\delta y = -\,\sqrt{2}\,\epsilon_i\xi^i - \sqrt{2}\,i\alpha\,\bar{\epsilon}_i\psi^i ,\qquad
    \delta \xi^i = \sqrt{2}\,i\bar{\epsilon}^i\left(\dot{y}-\alpha C\right) - \sqrt{2}\,\epsilon^i A,\qquad
    \delta A = -\,\sqrt{2}\,i\bar{\epsilon}_i\dot{\xi}^i,\nn
    &&\delta x = -\,\sqrt{2}\,\epsilon_i\psi^i ,\qquad
    \delta \psi^i = \sqrt{2}\,i\bar{\epsilon}^i\dot{x} - \sqrt{2}\,\epsilon^i C,\qquad
    \delta C = -\,\sqrt{2}\,i\bar{\epsilon}_i\dot{\psi}^i.
\eea
The components of $W$ have the standard transformations inherent to the chiral multiplet $({\bf 2, 4, 2})$,
while the transformations of the remaining fields $y$, $\xi^i$, $A$ acquire additional  pieces  involving the components of $W$ (they
are proportional to $\alpha\,$).

\subsection{Lagrangian}
The general kinetic Lagrangian is written as
\bea
    {\cal L}^{\rm kin}=\frac{1}{4}\int d^2\theta\,d^2\bar{\theta}\,f\left(V,\bar{V},W,\bar{W}\right),
\eea
where $f$ is just an arbitrary real function of superfields.
Like in \cite{Katona}, we can define six bilinear invariant kinetic Lagrangians:
\bea
    V\bar{V}, \quad W\bar{W}, \quad V\bar{W}+W\bar{V}, \quad i\left(V\bar{W}-W\bar{V}\right),\quad V^2+\bar{V}^2,
\quad i\left(V^2-\bar{V}^2\right).
\eea
Possible terms $VW$ and $\bar{W}\bar{V}$ do not contribute. Dependence on $\alpha$ remains only in the superfield $V$,
so only five out of six bilinear kinetic Lagrangians involve the parameter $\alpha$.

One can write a superpotential Lagrangian for the chiral superfield $W$ as
\bea
    {\cal L}^{\rm pot}_1 = \int d^2\theta\, {\cal F}\left(W\right)  + \int d^2\bar{\theta}\,\bar{\cal F}\,\big(\bar{W}\big)\,.
\eea
Another option is to write the following superpotential Lagrangian:
\bea
    {\cal L}^{\rm pot}_2 = \int d^2\theta\, h^\prime\left(W\right) V_0 + \int d^2\bar{\theta}\,\bar{h}^\prime\,\big(\bar{W}\big)\,\bar{V}_0\,.
\eea
The transformation property \eqref{V0tr} of $V_0$ allows to represent transformations of this term as
\bea
    \delta{\cal L}^{\rm pot}_2 = -\,i\alpha\bar{\epsilon}_k\int d^2\theta\,\frac{\partial}{\partial \theta_k}\,h\left(W\right) + \mbox{c.c.} = 0.
\eea
Above superpotential Lagrangians have no dependence on $\alpha$ since the chiral superfunction $V_0$ corresponds just to the limit $\alpha=0$ of $V$.

To make comparison with \cite{Katona}, we can consider two bilinear superpotential terms
\be
    \gamma_1\int d^2\theta\, V_0W + \bar{\gamma}_1\int d^2\bar{\theta}\,\bar{V}_0\bar W \quad \mbox{and} \quad
    \gamma_2\int d^2\theta\, W^2+ \bar{\gamma}_2\int d^2\bar{\theta}\,\bar{W}^2 .
\ee
Here, $\gamma_1$ and $\gamma_2$ are complex parameters of mass dimension. These bilinear superpotential terms in components generate
the so called ``Super-Zeeman''  invariant terms of ref.\cite{Katona}, one of them containing expressions proportional
to $\alpha$ and corresponding to a coupling to an external magnetic field. In \cite{Katona}, such terms  were also referred to
as Wess-Zumino type terms. This Wess-Zumino term can in fact be eliminated by redefining the fields (in the notations of ref. \cite{Katona}) as
\bea
F_4 \rightarrow F_4 - \alpha\dot{\phi}_6\,,\qquad \psi_1 \rightarrow \psi_1 +\alpha\psi_8\,,\qquad \psi_2 \rightarrow \psi_2 +\alpha\psi_7\,.
\eea
Then all ``Super-Zeeman'' invariant terms become independent of $\alpha$, and the same is true for our general superpotential Lagrangians.

In our notations, Wess-Zumino terms appear only after the elimination of auxiliary fields\footnote{Such a possibility was also mentioned
in \cite{Katona}.}.  An example of such a term is given in the next subsection.

\subsection{Free model}
As an  instructive example, let us consider the simple free Lagrangian given by
\bea
    {\cal L}^{\rm free}&=&\frac{1}{4}\int d^2\theta\,d^2\bar{\theta}\left[V\bar{V}+ \left(1 - \alpha^2\right)W\bar{W}\right]
+\frac{1}{4}\int d^2\theta\left(2\mu_1 V_0 + \mu_2 W\right)W\nn
    && + \,\frac{1}{4}\int d^2\bar{\theta}\left(2\mu_1\bar{V}_0 + \mu_2\bar{W}\right)\bar{W},\label{N4SF}
\eea
where $\mu_1$ and $\mu_2$ are real parameters of the mass dimension. The coefficient $\left(1 - \alpha^2\right)$
in front of $W\bar{W}$ was chosen to gain the correctly normalized kinetic terms in the off-shell Lagrangian:
\bea
    {\cal L}^{\rm free} &=& \dot{y}\dot{\bar{y}}+\dot{x}\dot{\bar{x}}+ \frac{i}{2}\left(\bar{\xi}_i\dot{\xi}^i
- \dot{\bar{\xi}}_i\xi^i\right)+\frac{i}{2}\left(\bar{\psi}_i\dot{\psi}^i - \dot{\bar{\psi}}_i\psi^i\right)
+ A\bar{A}+ \left(1 + \alpha^2\right) C\bar{C}
    - \alpha\left(C\dot{\bar{y}} + \bar{C}\dot{y}\right)\nn
    && +\,\mu_1 \left(C y + Ax + \bar{C}\bar{y} + \bar{A}\bar{x} + \xi^i\psi_i + \bar{\xi}_i\bar{\psi}^i\right)
+ \mu_2\left(C x + \bar{C}\bar{x} + \frac{1}{2}\,\psi^i\psi_i + \frac{1}{2}\,\bar{\psi}_i\bar{\psi}^i\right).
\eea
After eliminating the auxiliary fields $A$ and $C$ by their equations of motion,
\bea
    \left(1 + \alpha^2\right) C = \alpha\dot{y} - \mu_1\bar{y} - \mu_2\bar{x},\qquad
    A=-\,\mu_1 \bar{x},
\eea
and neglecting a total time-derivative, we obtain the on-shell Lagrangian
\bea
    {\cal L}^{\rm free} &=& \frac{\dot{y}\dot{\bar{y}}}{1 + \alpha^2}+\dot{x}\dot{\bar{x}}
+ \frac{i}{2}\left(\bar{\xi}_i\dot{\xi}^i - \dot{\bar{\xi}}_i\xi^i\right)+\frac{i}{2}\left(\bar{\psi}_i\dot{\psi}^i
- \dot{\bar{\psi}}_i\psi^i\right)
    +\frac{\alpha\mu_2\left( x\dot{y}+\bar{x}\dot{\bar y}\right)}{1 + \alpha^2}
    -\left(\mu_1\right)^2 x\bar{x}\nn
    &&-\,\frac{\left(\mu_1\bar{y} + \mu_2\bar{x}\right)\left(\mu_1 y + \mu_2 x\right)}{1 + \alpha^2}
+ \mu_1\left(\xi^i\psi_i + \bar{\xi}_i\bar{\psi}^i\right)
    +\frac{\mu_2}{2}\left(\psi^i\psi_i + \bar{\psi}_i\bar{\psi}^i\right).\label{N4free}
\eea
The relevant on-shell transformations are given by
\bea
    &&\delta y = -\,\sqrt{2}\,\epsilon_i\xi^i - \sqrt{2}\,i\alpha\,\bar{\epsilon}_i\psi^i ,\quad
    \delta \xi^i = \frac{\sqrt{2}\,i\bar{\epsilon}^i}{1 + \alpha^2}\left[\dot{y}+\alpha\left(\mu_1\bar{y}
+ \mu_2\bar{x}\right)\right] + \sqrt{2}\,\mu_1\,\epsilon^i \bar{x},\nn
    &&\delta x = -\,\sqrt{2}\,\epsilon_i\psi^i ,\qquad
    \delta \psi^i = \sqrt{2}\,i\bar{\epsilon}^i\dot{x} - \frac{\sqrt{2}\,\epsilon^i}{1 + \alpha^2}\left(\alpha\dot{y}
- \mu_1\bar{y} - \mu_2\bar{x}\right) .\label{onshelltr}
\eea
The on-shell Lagrangian \eqref{N4free} contains the Wess-Zumino type term describing an interaction between
two chiral ${\cal N}=4$ multiplets ${\bf (2,4,2)}$:
\bea
    \sim\frac{\alpha\mu_2\left(x\dot{y}+\bar{x}\dot{\bar y}\right)}{1 + \alpha^2}.
\eea
This term matches with the statement of \cite{Katona} that the elimination of auxiliary fields induces additional
terms which can be treated as a coupling to an external magnetic field.

When $\mu_1=\mu_2=0$, we can make rescaling $y\rightarrow \sqrt{1+\alpha^2}\,y$ in the Lagrangian \eqref{N4free}
and obtain the $\alpha$-independent Lagrangian
\bea
    {\cal L}^{\rm free}\left.\right|_{\mu_1=\mu_2=0} = \dot{y}\dot{\bar{y}}+\dot{x}\dot{\bar{x}}+ \frac{i}{2}\left(\bar{\xi}_i\dot{\xi}^i
- \dot{\bar{\xi}}_i\xi^i\right)+\frac{i}{2}\left(\bar{\psi}_i\dot{\psi}^i - \dot{\bar{\psi}}_i\psi^i\right).\label{N4kin}
\eea
The transformations \eqref{onshelltr} become
\bea
    &&\delta y =\frac{1}{\sqrt{1+\alpha^2}}\left[ -\,\sqrt{2}\,\epsilon_i\xi^i - \sqrt{2}\,i\alpha\,\bar{\epsilon}_i\psi^i \right],\qquad
    \delta \xi^i = \frac{\sqrt{2}\,i\bar{\epsilon}^i\dot{y}}{\sqrt{1+\alpha^2}}\,,\nn
    &&\delta x = -\,\sqrt{2}\,\epsilon_i\psi^i ,\qquad
    \delta \psi^i = \sqrt{2}\,i\bar{\epsilon}^i\dot{x} - \frac{\sqrt{2}\,\alpha\,\epsilon^i\dot{y} }{\sqrt{1+\alpha^2}}\,.\label{not0}
\eea
The Lagrangian \eqref{N4kin} is thus invariant under supersymmetry transformations with various parameters $\alpha$,
since it has no dependence on $\alpha$. For instance, it is invariant under the undeformed $\alpha=0$ transformations
\bea
    \delta y =-\,\sqrt{2}\,\eta_i\xi^i ,\qquad
    \delta \xi^i = \sqrt{2}\,i\bar{\eta}^i\dot{y},\qquad
    \delta x = -\,\sqrt{2}\,\eta_i\psi^i ,\qquad
    \delta \psi^i = \sqrt{2}\,i\bar{\eta}^i\dot{x}.\label{0}
\eea
Their closure with \eqref{not0} yields additional bosonic transformations
\bea
    \delta y = a\dot{x},\qquad  \delta x = \bar{a}\dot{y},
\eea
which leave  the Lagrangian \eqref{N4kin} invariant and commute (on-shell) with the supersymmetric transformations
\eqref{not0} for any $\alpha$. It would be interesting to see whether a similar phenomenon takes place in the interaction
case too.

\section{Summary and outlook}

We have shown how long ${\cal N}=2$,  $d=1$ multiplets can be embedded into $SU(2|1)$ chiral multiplets in the framework
of $SU(2|1)$ supersymmetric mechanics. They naturally appear in $SU(2|1)$ mechanics of chiral multiplets,
when the chiral superfield $\Phi_A$ carries some external index  $A$ with respect to the subgroup  $SU(2)$
of the supergroup $SU(2|1)$. We studied this multiplet in the framework of ${\cal N}=2$ superspace and constructed
its general superfield action.

We considered the long ${\cal N}=4$ multiplet \cite{Katona} within the standard ${\cal N}=4$ superspace.
Defining and solving the constraint \eqref{alpha}, we obtained the superfields $V$ and $W$
describing the long multiplet ${\bf (4, 8, 4)}_{\rm l}$ and constructed general superfield Lagrangians
consisting of the kinetic (sigma-model type) and superpotential Lagrangians. We considered
the free Lagrangian \eqref{N4SF}, where the superpotential term $\sim\mu_2$ is responsible
for appearing Wess-Zumino type term in the on-shell Lagrangian \eqref{N4free}.

In conclusion, we outline some further possible lines of investigation.
\begin{itemize}
\item Quantization of the model \eqref{N4free} and construction of the Hilbert space of wave functions.
\item Study of some other generalizations of long multiplets to the standard flat  ${\cal N}=4$,  $d=1$ supersymmetry \cite{GonTo1,GonTo}.
\item Generalizing the constraints \eqref{alpha} to the $SU(2|1)$ covariantized constraints:
\bea
    \bar{{\cal D}}_k V = i\alpha {\cal D}_k W,\qquad \bar{{\cal D}}_k W = 0.
\eea
Such a generalization  is possible for the second type of the $SU(2|1)$ chirality \cite{SKO}, when the spinor derivatives
are inert under the induced $U(1)$ transformations.
\item Answering the question whether it is possible to find out $d>1$ analogs of long multiplets.
\end{itemize}

\ack
E.I. and S.S. thank the organizers of the conference ISQS24 for the kind hospitality in Prague. S.S. acknowledges support
from the Votruba-Blokhintsev program. The work of E.I. and S.S. was partially supported by the Russian Science Foundation, Grant No.~16-12-10306.
A.R.M. thanks the Directorate of BLTP and JINR for the opportunity to work in a scientific atmosphere.

\end{document}